\documentclass[10pt,journal,compsoc]{IEEEtran}

\usepackage[T1]{fontenc}
\usepackage[utf8]{inputenc}
\usepackage{placeins}
\usepackage{microtype}

\ifCLASSOPTIONcompsoc
  \usepackage[nocompress]{cite}
\else
  \usepackage{cite}
\fi
\ifCLASSINFOpdf
  \usepackage[pdftex]{graphicx}
\else
\fi
\usepackage{amsmath}
\usepackage{amssymb}
\usepackage{xfrac}

\usepackage{array}

\usepackage{tabularx}
\usepackage{booktabs}
\usepackage{threeparttable}

\ifCLASSOPTIONcompsoc
 \usepackage[caption=false,font=footnotesize,labelfont=sf,textfont=sf]{subfig}
\else
 \usepackage[caption=false,font=footnotesize]{subfig}
\fi

\usepackage{stfloats}
\fnbelowfloat
\usepackage[colorlinks=true,allcolors=blue]{hyperref}

\begin{document}
%
\title{Multimodal Classification of Stressful \\ Environments in Visually Impaired Mobility \\ Using EEG and Peripheral Biosignals}
%
%
%

\author{Charalampos~Saitis
        and~Kyriaki~Kalimeri,~\IEEEmembership{Member,~IEEE}
\IEEEcompsocitemizethanks{\IEEEcompsocthanksitem C.~Saitis is with the Audio Communication Group,  Technische Universität Berlin, Germany. E-mail: \href{mailto:charalampos.saitis@campus.tu-berlin.de}{charalampos.saitis@campus.tu-berlin.de}.
\IEEEcompsocthanksitem K. Kalimeri is with the Institute for Scientific Interchange (ISI Foundation), Turin, Italy. E-mail: \href{mailto:kyriaki.kalimeri@isi.it}{kalimeri@ieee.org}.}
\thanks{Manuscript received X Mon.~Year; revised X Mon.~Year; accepted X Mon.~Year. Date of publication X Mon.~Year; date of current version X Mon.~Year.\protect\\
Recommended for acceptance by John Snow.\protect\\
For information on obtaining reprints of this article, please send e-mail to:
reprints@ieee.org, and reference the Digital Object Identifier below.\protect\\
Digital Object Identifier no. XXXXXXX}}

%
%

\markboth{IEEE Transactions on Affective Computing,~Vol.~XX, No.~X, Month~Year}%
{Shell \MakeLowercase{\textit{et al.}}: Bare Demo of IEEEtran.cls for Computer Society Journals}
%



\IEEEtitleabstractindextext{%
\begin{abstract}
In this study, we aim to better understand the cognitive-emotional experience of visually impaired people when navigating in unfamiliar urban environments, both outdoor and indoor.
We propose a multimodal framework based on random forest classifiers, which predict the actual environment among predefined generic classes of urban settings, inferring on real-time, non-invasive, ambulatory monitoring of brain and peripheral biosignals.  
Model performance reached 93\% for the outdoor and 87\% for the indoor environments (expressed in weighted AUROC), demonstrating the potential of the approach. 
Estimating the density distributions of the most predictive biomarkers, we present a series of geographic and temporal visualizations depicting the environmental contexts in which the most intense affective and cognitive reactions take place.
A linear mixed model analysis revealed significant differences between categories of vision impairment, but not between normal and impaired vision. Despite the limited size of our cohort, these findings pave the way to emotionally intelligent mobility-enhancing systems, capable of implicit adaptation not only to changing environments but also to shifts in the affective state of the user in relation to different environmental and situational factors.
\end{abstract}

}

\maketitle

\IEEEdisplaynontitleabstractindextext

%
\IEEEpeerreviewmaketitle

\IEEEraisesectionheading{\section{Introduction}\label{sec:introduction}}

%
%
%
%
\IEEEPARstart{M}{obility} in indoor and outdoor environments can be a challenging and emotionally stressful task for visually impaired people (VIP), especially when navigating in unfamiliar sites. Despite an increasing number of assistive technologies that help individuals with sight loss to augment their spatial awareness and wayfinding abilities when in move, very few systems provide a high degree of independence beyond known environments that would allow VIP to significantly achieve mobility and integrate into everyday active life \cite{giudice2008,marston2003}. Placing the visually impaired in the center of attention and exploiting recent developments in pervasive physiological sensing for affective computing, two mobility ``in the wild'' studies were designed to better understand how people with sight loss perceive and interact with the urban and indoor space as manifested in their management of cognitive load and stress. 

Orientation and mobility (O\&M) in humans heavily relies on sight, which provides instantaneous, effortless access to anticipatory (e.g., stairs, turns, signs) and proactive (e.g., moving people, poles) information at various distances simultaneously \cite{millar1994}. Visually impaired pedestrians learn to obtain critical environmental information primarily through touch (sensing the ground surface with a white cane) and hearing (identifying and localizing events and landmarks through sound). Mobility challenges can be summarized in four main problems: avoiding obstacles (e.g., people moving or standing in the way, pillars, tree branches, doors opening outwards, improperly parked cars); detecting ground level changes (e.g., stairs, ramps, pavement edge); negotiating street crossings (e.g., lack of curbs, traffic lights, or sound signaling); finding entrance/exit points (e.g., automated doors, elevators); and adapting to light variation (e.g., abrupt changes between different environments) \cite{geruschat2010,quinones2011}. 

Although these problems generally diminish with increased experience of an environment (e.g., own living or working space, route from home to work), they still make traveling in unfamiliar settings particularly challenging, often preventing VIP from going outdoors or visiting new sites altogether. The limitations and dependence on others in daily living and mobility often have profound consequences for the psychological health of VIP, generating increased anxiety and motivating social isolation and depression \cite{Wallhagen2001,Rees2010}. An increasing number of navigation and access technologies has removed many barriers to independent wayfinding for VIP, greatly advancing their personal and social well-being. However, the degree of independence offered by these technologies can be limited when the user encounters unfamiliar situations that stress and inhibit them \cite{giudice2008}. A less than optimal presentation of information may cause unnecessary mental burden, increasing emotional stress through imposing cognitive load on working memory \cite{Sweller1994}.

Despite a significant amount of research on understanding the perceptual and neurocognitive mechanisms by which people with sight loss access and process wayfinding information \cite{cattaneo2008}, there is still little practical knowledge of how the management of cognitive load and psychological stress relates to the wayfinding process itself. This is a critical aspect of human-computer interaction that has only recently been considered essential in designing \emph{emotionally intelligent} mobility-enhancing systems that are capable of implicitly adapting not only to changing environments but also to shifts in the user's affective experience in relation to environmental factors \cite{welsh2010}. For example, to activate additional information if a visually impaired traveler feels unsafe and stressed due to an insufficient representation of the surroundings of the system without increasing the mental workload of the user, or to reduce the amount of channeled information if a VIP feels relaxed and confident in a certain environment, or to evaluate changing priorities.  

\begin{figure*}[!t]
\centering
\includegraphics[width=\textwidth]{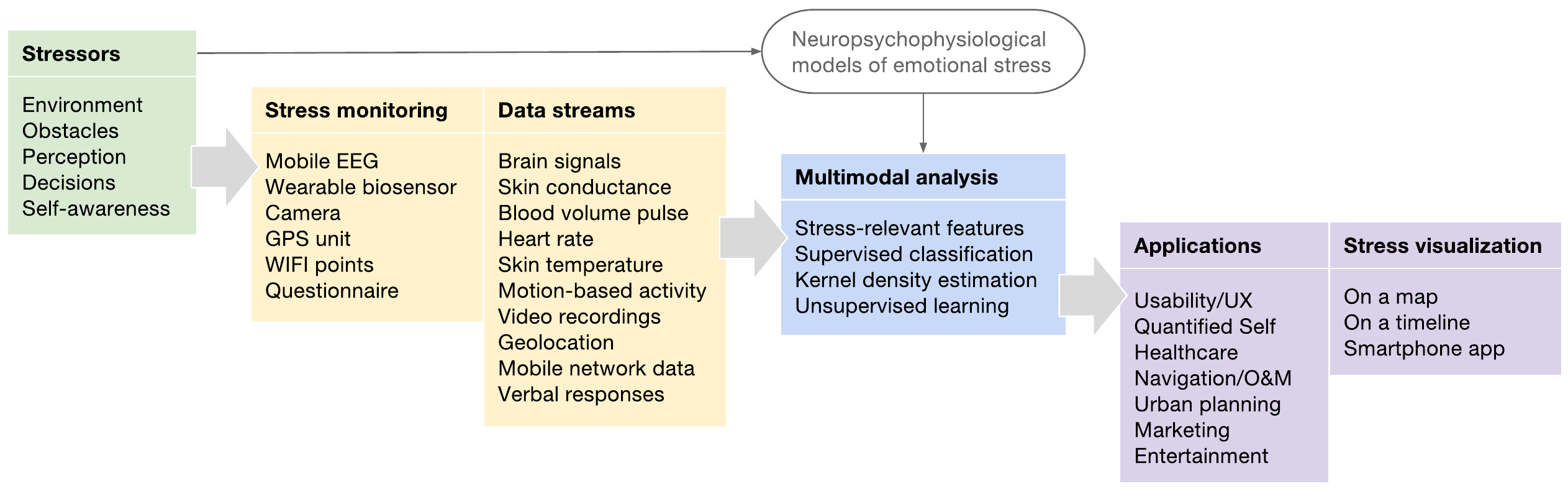}
\caption{Multimodal biosignal data capture and analysis framework for detecting stress during mobility tasks with visually impaired people. Adapted from Fig.~1 in \cite{kanjo2015}.}
\label{fig:framework}
\end{figure*}

One way of detecting emotion and psychological stress is through identifying patterns in the central and peripheral physiological modalities, the most common being electrodermal activity (EDA), cardiovascular activity, and electroencephalography (EEG). Electrodermal activity is a well-known indicator of physiological arousal and stress activation in affective computing \cite{Picard2005,setz2010}. It is more sensitive to emotion-related variations in arousal as opposed to physical stressors, which can be better reflected in measurements of heart rate (HR). Blood volume pulse (BVP) patterns can also reflect transient processes in arousal and cognitions \cite{peper2007}. Two outdoor mobility studies in the early 1970s suggested that some form of psychological rather than physical stress is responsible for increased HR in visually impaired versus sighted pedestrians \cite{peake1971,wycherley1970}. However, certain mobility tasks (e.g., stair climbing) may result in an interactive effect of  psychological stress and momentary physical workload, thus cardiovascular measures may be less suitable than EDA.

Electroencephalography, on the other hand, can provide neurophysiological markers of cognitive-emotional processes induced by stress due to imposed cognitive load, indicated by changes in brain activity. The latter is characterized by rhythmic patterns across distinct frequency bands, the definition of which can vary somewhat among studies. Hereafter we consider six EEG bands, namely delta (0.5--4 Hz), theta (4--7 Hz), alpha-1 (7--10 Hz), alpha-2 (10--13 Hz), beta (13--30 Hz), and gamma (30--60 Hz). Gamma waves are thought to be involved in higher cognitive functions such as multimodal processing or object representation \cite{keil1999}. Beta waves are associated with psychological and physical stress \cite{jena2015}, whereas theta and alpha-1 frequencies reflect response inhibition and attentional demands such as phasic alertness \cite{ray1985}. Alpha-2 is related to task performance in terms of speed, relevance, and difficulty \cite{Gevins1997,Klimesch1999}. Reduced activity at alpha frequencies has been repeatedly associated with increasing cognitive load in a variety of task demands (see \cite{Antonenko2010} for a review). Further evidence suggests that asymmetry in frontal alpha band power varies dependent on affective disposition and engagement, with more activity in left alpha indicating positive approach/motivation and emotions, whereas increased right frontal alpha showed withdrawal/avoidance and negative emotions \cite{harmonjones2010}. EEG delta activity has been reported to indicate attention to internal processing during performance of mental tasks \cite{harmony1996}.

In recent years, the advent of ubiquitous mobile and sensing technologies, consumer brain-computer interfaces (BCI), and the quantified self movement has driven the development of wireless wearable multi-sensor systems (from devices to smartphone apps) for easy and reliable automatic collection of brain and peripheral biosignal data streams, making it possible to monitor human affective states in virtually any real-world situation \cite{Debener2012,kanjo2015}. Massot and colleagues \cite{massot2012} used a custom mobile biosensor to collect EDA from 27 blind pedestrians as they walked through urban environments of varying complexity. Examination of arousal-relevant EDA features showed that VIP experience increased psychological stress when walking on busy shopping streets, passing through large open areas, and crossing junctions. In another study \cite{mavros2015}, analyses of EEG signals recorded from 12 VIP during outdoor travel using a commercial BCI headset \cite{mavros2015} further indicated that busy streets, open spaces, and street crossings induce larger cognitive engagement than quieter and less complex urban settings.

Expanding on previous work by the authors and colleagues \cite{saitis16a,Kalimeri2016,saitis2018cogload}, this paper presents a multimodal framework to automatic inference of stressful environmental conditions affecting visually impaired mobility based on ambulatory monitoring and multimodal fusion of EEG, EDA, and BVP signals, taking advantage of their inherent and complementary properties (Fig. \ref{fig:framework}). The goal of the research was twofold: to discover biomarkers that can be used to detect shifts in emotional stress and cognitive load between different settings and situations, and to develop and understanding of which environmental factors increase cognitive load and stress during visually impaired mobility. Such knowledge can help design \emph{emotionally intelligent} O\&M systems, which are capable of implicitly adapting not only to changing environments but also to shifts in the internal experience of the user in relation to environmental factors. The proposed framework thus differs fundamentally from context-based approaches to environment recognition, for example, GPS-based geolocation. While the latter allow a certain degree of independence for VIP, identifying dynamic stressors in different or the same environments can lead to even more independent mobility systems that recognize, interpret, and adapt to the affective states of the user.

Using state-of-the-art portable sensor devices, EEG, EDA, and BVP signals were collected from a group of VIP and from two normally sighted individuals and as they walked through outdoor and indoor environments of varying complexity and difficulties. A number of multimodal features ranging from low-level signal descriptors to indexes of higher cognitive and emotional functions were extracted and used in unimodal and multimodal classification experiments. While the relationship between unimodal biosignals and psychological arousal has been studied extensively, the detection of stress from fusing multimodal biosignal streams has not been comparatively investigated. To better understand the relationship of stress biomarkers with the environmental and situational factors that evoke them, the most predictive features were examined in relation to variables such as type of environment/situation, amount of vision loss, and impaired versus normal sight with a linear mixed model analysis. A technique for visualizing geographical and temporal density distributions of biomarkers using weighted kernel density estimation \cite{wand1995} and dynamic time warping \cite{giorgino2009} was developed.

\section{Materials and Methods}


\begin{figure}[!t]
\centering
\includegraphics[width=0.8\columnwidth]{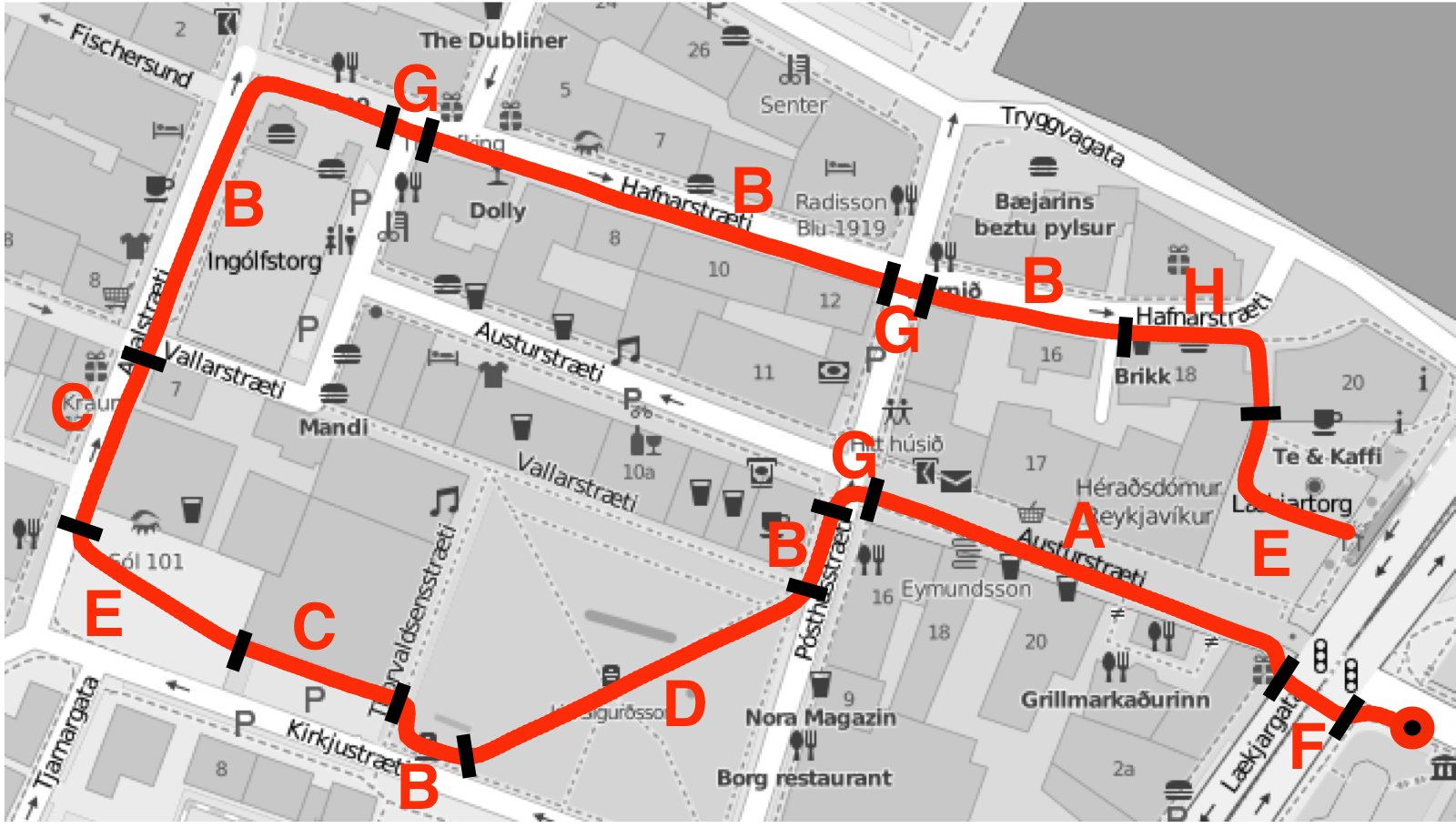}
\caption[Outdoor route. Letters depict the different urban environments reported in Table \ref{tab:environments}; black bars indicate where they start/end; the red-black dot shows the starting point of the walk.]{Outdoor route. Letters depict the different urban scenes reported in Table \ref{tab:environments}; black bars indicate where they start/end; the red-black dot shows the starting point of the walk. Map image provided by the OpenStreetMap (OSM) collaborative project (\url{https://www.openstreetmap.org/}).}
\label{fig:route_out}
\end{figure}

\begin{figure}[!t]
\centering
\includegraphics[width=\columnwidth]{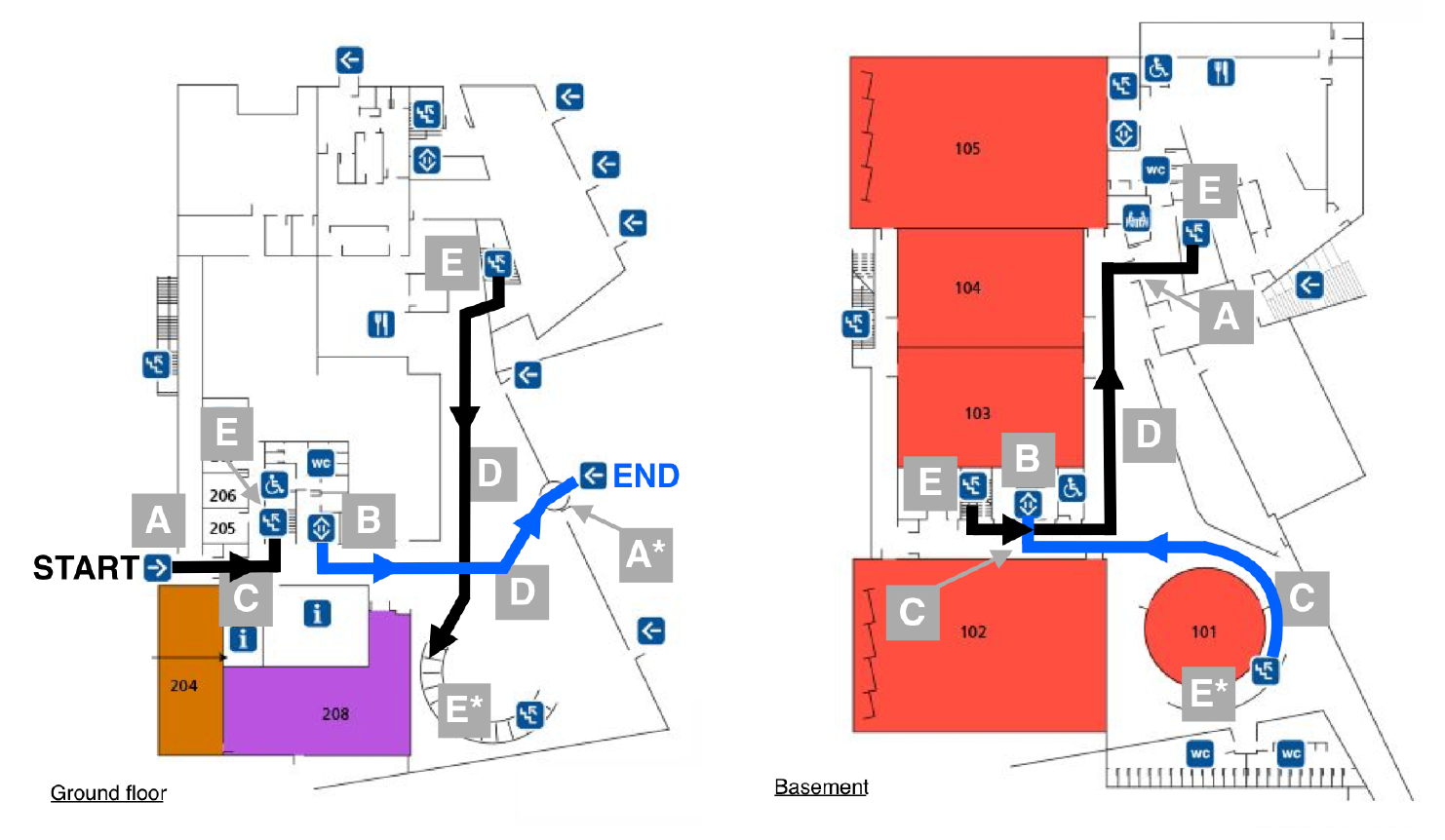}
\caption{Indoor study: layout of the site and charted route. Letters depict the indoor environments reported in Table \ref{tab:environments}, with A$^*$ indicating the rotating door and E$^*$ the large spiral stairs.}
\label{fig:route_in}
\end{figure}

\subsection{Participants}

A total of ten healthy visually impaired adults with different degrees of sight loss participated in the two mobility studies (6 female; average age = 41 yrs, range = 22--53 yrs). One participant was fully blind, three had visual acuity less than 2\%, four had visual acuity less than 5\%, and two had visual acuity between 5\% and 10\%.\footnote{Based on the classification of visual impairment by the World Health Organization: \url{http://apps.who.int/classifications/apps/icd/icd10online2003/fr-icd.htm?gh53.htm+}.} Eight of them were congenitally or early blind (first 2--3 yrs of life) and two had become blind later in life (generally after the age of 3). To help make the VIP feel comfortable and safe, they were encouraged to walk as usual using their white canes if they wished so, and were accompanied by their familiar O\&M instructor. Participants were instructed to avoid smoking normal or e-cigarettes and consuming caffeine or sugar (e.g., coffee, cola, chocolate) approximately one hour prior to the walk. Recruitment was based on volunteering and all VIP were capable of giving free and informed consent. The study was approved by the National Bioethics Committee of Iceland. All data was anonymized before analysis.

All visually impaired participants actively experience indoor environments other than where they reside on a daily basis: four work full-time, three are part-time employees, and three attend educational or vocational establishments. All VIP reported traveling alone outdoors on an ``almost daily'' basis, but six of them would not feel confident enough to do so in unfamiliar spaces and routes. Two participants, who reported regular use of a white cane when mobile, felt safe enough to walk without any aid. When asked to describe their feelings regarding the ease of mobility over the previous year, four VIP believed that it has not changed, an equal number thought that it has become easier, while two considered it to have become less easy. Eight of the participants walked both the outdoor and indoor routes, one took part only in the outdoor study, and two completed only the indoor task. 

Two healthy normally sighted individuals (1 female; age = 31 and 40 yrs) were further recruited. They walked only the outdoor route and their data were used only in the linear mixed model analysis (Sec.~\ref{sec:lma}). 

\begin{table*}[!t]
\centering
\caption{Descriptions of Mobility Environments}
\renewcommand{\arraystretch}{1.1}
\begin{tabularx}{\textwidth}{@{\extracolsep{\fill}}l >{\raggedright\arraybackslash}p{5.5cm}  >{\raggedright\arraybackslash}p{11.5cm} }
\toprule
\textbf{ID} & \textbf{Description}  & \textbf{Challenges} \\
\midrule
\multicolumn{3}{@{}l}{\emph{Outdoor Route}} \\
A & Shopping street       & People, ads, chairs, tables, poles,  ramps   \\
B & Small street          & People, poles, ads,  ramps,  blocked passageway      \\
C & Narrow alley          & People, chairs, tables, street ads, trash bins, flower planters, stairs going down, parked cars \\
D & Urban park            & People, tree branches, poles, flower planters, blocked passageway \\
E & Open space            & People, flower planters, stairs going up, blocked passageway      \\
F & Crossing main road with traffic lights   & People       \\
G & Crossing small street without traffic lights & People, uneven pavement, detecting edges  \\
H & Construction alley    & People, ramps, construction  \\
\midrule
\multicolumn{3}{@{}l}{\emph{Indoor Route}} \\
A & Entering through automated doors (two hinged and one rotating) &
Finding the pushbutton (hinged doors only), finding where and when to enter the rotating door, other people going through the door at the same time \\ 
B  & Using an elevator to move between floors &
Finding the pushbuttons (calling the elevator, selecting floor), other people exiting/entering\\ 
C &  Walking along a narrow corridor & 
Moving people, noise, classroom doors opening suddenly  \\ 
D & Moving across an open space &
Moving people, standing people, tables, chairs, trash bins, pillars, people talking, loud noises \\ 
E & Using stairs to move between floors &
People using the stairs in the opposite direction, walk down a large spiral flight of stairs \\ 
\bottomrule
\end{tabularx}
\label{tab:environments}
\end{table*}

\subsection{Mobility Environments}

The outdoor and indoor routes were planned with the assistance of caretakers and O\&M instructors to take the VIP through circumstances of varying complexity and difficulty, where different levels of stress were likely to occur.

\emph{Outdoor~Route}. The charted course was located in the part of Reykjavik's city center between the City Hall and the port, which consists of smaller and larger streets, narrower and wider sidewalks, street crossings with and without traffic lights, as well a number of open spaces. Accordingly, the route was divided into seventeen parts grouped in eight distinct urban environments (see Fig.~\ref{fig:route_out} and upper part of Table \ref{tab:environments}). These were defined so as to cluster environmental and situational factors expected to elicit similar affective reactions. For example, participants had to walk on a busy shopping street (environment A1), pass through an urban park-square (D1), and cross a major junction (F1). The route was approximately 1 km long and took on average 13 min 44 sec to walk (range = 9--19 min). 

\emph{Indoor~Route}. The H\'{a}sk\'{o}latorg building of the University of Iceland campus in Reykjavik houses various service units for students, a bookstore, two restaurants, classrooms, and reading rooms. As such, it provided both uncomplicated and sufficiently complex indoor scenes for the purposes of our study. The charted route linked the entrance at the back of the building (START) to the main entrance at its front (END) and comprised five distinct environments representable of a variety of indoor mobility challenges (see Fig.~\ref{fig:route_in} and lower part of Table \ref{tab:environments}). Indicatively, participants had to enter through automated doors (scenes A2 and A2$^*$), use an elevator (B2), move across a busy open space (D2---main entrance hall), and walk down a large spiral staircase (E2$^*$). The route was approximately 200 meters in length and took on average 5 minutes to walk (range = 4--8 minutes). 



\subsection{Multimodal Biosignals}


\emph{EEG~Registration}. Brain electrical signals were recorded using the Emotiv EPOC\verb!+! (\url{http://emotiv.com/epoc/}), a mobile headset with 16 dry electrodes registering over the 10-20 system locations AF3, F7, F3, FC5, T7, P3 (CMS), P7, O1, O2, P8, P4 (DRL), T8, FC6, F4, F8, and FC4 (sampling rate $f_s = 128$ Hz). Given the practical constraints involved in monitoring brain electrical activity in the wild, EPOC\verb!+! was chosen because it provides a good compromise between performance (i.e., number of channels and scientific validity of the acquired EEG signals) and usability (i.e., portability, preparation time and user comfort) with respect to other commercial wireless EEG systems \cite{badcock2013,hairston2014,Debener2012,ekandem2012}. 

\emph{EDA/BVP~Registration}. Along with the Emotiv headset, participants were asked to wear the Empatica E4 wristband (\url{https://www.empatica.com/e4-wristband}) \cite{garbarino2014}. E4 measures EDA as skin conductance through 2 ventral (inner) wrist electrodes ($f_s = 4$ Hz) and BVP through a dorsal (outer) wrist photoplethysmography (PPG) sensor ($f_s = 64$ Hz). E4 further reports HR, extracted on board from BVP interbeat intervals. The wristband also includes an infrared thermopile sensor and a 3-axis accelerometer. E4 is currently the only commercial multi-sensor device developed based on extended scientific research in the areas of psychophysiology and affective computing. Additionally, it has a cable-free, watch-like design, which makes it easier and more aesthetically pleasing to wear, and thus better fitted to use in the wild compared to other wearable biosignal devices. Participants were asked to wear the wristband on the non-dominant hand to minimize motion artifacts related to handling the white cane \cite{boucsein2012book}.  

\subsection{General Procedure}

Participants walked the outdoor route twice and the indoor route three times for training purposes. In both studies directions were only provided during the first walk to help the VIP familiarize with the route. They were instructed to avoid unnecessary head movements and hand gestures as well as talking to their O\&M instructor unless there was an emergency. Video and audio were registered by means of a smartphone camera to facilitate data annotation (observing behaviors across the different environments and situations) and synchronization (start/end of walk, environments, and obstacles). In the outdoor study, GPS coordinates were additionally logged using a Garmin GPSMAP-64s unit at a rate of 1 registration per second. Upon completing the last walk, participants were asked to describe stressful moments they experienced along the route.


\subsection{Feature Extraction}\label{sec:features}

\emph{EEG~Features}. The Emotiv EPOC\verb!+! system involves a number of internal signal conditioning steps. Analogue signals are first high-pass filtered with a 0.16 Hz cut-off, pre-amplified, low-pass filtered with an 83 Hz cut-off, and sampled at 2048 Hz. Digital signals are then notch-filtered at 50/60 Hz and down-sampled to 128 Hz prior to transmission. For approximately half of the participants, EEG data obtained from the headset was first time-domain interpolated using the Fast Fourier Transform (FFT) to account for missing samples due to connectivity issues. Subsequently, all signals were baseline-normalized. For visually impaired participants, this involved subtracting for each individual and for each channel the mean of resting state registrations obtained during a series of laboratory studies with the same participants \cite{spagnol16a}. For normally sighted participants, for whom no resting state EEG data were available due to technical issues, the mean signal value was instead subtracted for each individual and for each channel. Finally, min-max scaling was applied to reduce inter-individual variance. 

A number of features related to signal power and complexity were extracted using the PyEEG open source Python module \cite{pyeeg}. For each of the 14 EEG channels, we computed the relative spectral power \cite{quiroga1997} in the delta (0.5--4 Hz), theta (4--7 Hz), alpha-1 (7--10 Hz), alpha-2 (10--13 Hz), beta (13--30 Hz), and gamma (30--60 Hz) bands using the Power Spectral Intensity (PSI) and Relative Intensity Ratio (RIR) functions:
\begin{align*}
\text{PSI}_k & = \displaystyle\sum_{i = |N(f_k / f_s)|}^{|N(f_{k+1} / f_s)|}{|X_i|} \quad \text{and} \\[1ex]
\text{RIR}_k & = \frac{\text{PSI}_k}{\sum_{j=1}^{K-1}{\text{PSI}_j}}, \quad k = 1, 2, \ldots, K-1 
\end{align*}  
where $f_s$ is the sampling rate, $N$ is the time series length, $|X_1, X_2, \ldots, X_N|$ is the FFT of the series, and $K$ is the total number of bands. Only RIR was considered in the subsequent multimodal analyses. We then calculated Spectral Entropy, which yields the entropy of the power spectrum (across bands) \cite{inouye1991}. It is defined as 
\[
\text{H} = - \frac{1}{log\;(K)} \displaystyle\sum_{i=1}^{K} \text{RIR}_i\;log\;(\text{RIR}_i)
\]
where RIR is as above. Finally we computed SVD Entropy, which measures entropy over the spectrum of eigenvalues in a singular value decomposition (SVD) of an embedding matrix formed by consecutive delay vectors extracted from the signal \cite{roberts99}. It is defined as 
\[
\text{H}_{\text{SVD}} = - \displaystyle\sum_{i=1}^{M} \overline{\sigma_i} log_2 \overline{\sigma_i}
\]
where $M$ is the number of singular values and $\overline{\sigma_1}, \overline{\sigma_2}, \ldots, \overline{\sigma_M}$ are normalized singular values such that $\overline{\sigma_i} = \sigma_i / \sum_{j=1}^{M}\sigma_j$. PyEEG uses an embedding dimension of $d_E = 20$ and delay $\tau = 2$. 

We further computed, for each electrode and for each frequency band, the event-related (de-) synchronization (ERD/ERS), which reflects the decrease (event-related desynchronization; ERD) or increase (event-related synchronization; ERS) in-band power while performing a task relative to a reference baseline without any task demands, in our case the resting state \cite{Pfurtscheller1999}. It is defined as 
\[
\text{ERD/ERS}_k = \frac{\text{resting RIR}_k - \text{RIR}_k}{\text{resting RIR}_k} *100
\]
where RIR and $k = 1, 2, \ldots, K$ are as previously. Positive ERD/ERS values indicate a decrease in band power (ERD), while negative values indicate an increase (ERS). 

Lastly, a Frontal Asymmetry Index (FAI) was computed by subtracting the log-transformed alpha power in the F3 channel (left frontal) from the log-transformed alpha power in the F4 channel (right frontal) \cite{Allen2004}. 
\[
\text{FAI} = log\bigg(\frac{\text{RIR}(\text{F4})_{alpha}}{\text{RIR}(\text{F3})_{alpha}}\bigg)
\]
RIR is defined as above. Because increased brain activity suppresses alpha waves, higher values on this index reflect relatively higher left activity (i.e., lower left alpha power) and thus positive feelings and higher engagement. The FAI index was calculated for both the alpha-1 (lower alpha) and alpha-2 (upper alpha) bands.
In total 198 features were extracted from each individual EEG recording, using time windows equal to one second of the continuous signal.

\emph{EDA~Features}. A measurement of skin conductance is traditionally characterized by two types of behavior: short-lasting phasic responses (can be thought of as rapidly changing peaks in EDA) and long-term tonic level in the absence of phasic responses (can be thought of as the underlying slow-changing level of EDA). In terms of physiology, a skin conductance response (SCR) is a sudden rise in the electrical conductance of the skin due to secretion from the skin's sweat glands (sweat contains electrolytes) in response to sympathetic nervous activation. Another characteristic of the signal is the superposition of subsequent SCRs (i.e., one SCR emerges on top of the preceding one), typically observed in states of high arousal. 

Skin conductance data obtained from the E4 was first low-pass filtered (1st order Butterworth, $f_c = 0.6$ Hz) to remove steep peaks stemming from artifacts and subsequently min-max normalized to reduce inter-individual variance \cite{cacioppo}. Conditioned SC signals were then decomposed into two continuous components of phasic and tonic EDA by means of deconvolution using the biexponential function as impulse response and estimating tonic activity, and implicitly phasic activity, through inter-impulse fitting of the deconvolved SC data \cite{benedek2010}. This decomposition and subsequent extraction of tonic and phasic EDA features was performed in Ledalab, a Matlab based toolbox (\url{http://www.ledalab.de/}). Six features were extracted: mean tonic EDA and the number of ``spontaneous'' SCRs (i.e., phasic changes not traceable to specific stimulation), which are known to be particularly suitable for longitudinal monitoring of emotional stress-elicited EDA (i.e., tonic arousal); sum of amplitudes of registered SCRs and average, maximum, and cumulative phasic EDA, which provide varying indicators of instantaneous phasic arousal \cite{boucsein2012book}.
We also used the EDA signal directly as reported from the E4 wristband, applying only min-max scaling to lessen inter-individual variation \cite{cacioppo}.   

\emph{BVP~Features}. To index cardiovascular activity, we used the BVP and HR data as streamed by the Empatica E4 wristband. HR is derived from BVP interbeat intervals. The raw BVP signal is preprocessed on board using a proprietary motion artifact removal technique \cite{garbarino2014}. No further conditioning was implemented to either of the signals. However, given that cardiovascular markers can be highly dependent on physical activity (e.g., when climbing stairs), BVP and HR were min-max normalized prior to analysis.

\subsection{Classification Analysis}

In order to identify automatically the affective meaning of an urban space based on biosignals recorded from VIP walking through it, we postulated the study as a supervised classification process. A widely-used ensemble learning method for classification was employed, namely Random Forest (RF) classifier \cite{breiman2001}, selected due to its ability to deal with imbalanced classes as well as because it provides a straightforward assessment of the variable importances. For each of the distinct environments described in Table \ref{tab:environments}, each time point of the corresponding biosignal data was annotated based on a binary schema per second, where ``1'' signaled the presence of the participant in the given environment at the given time point and ``0'' otherwise. 

A series of experiments were designed to assess and compare the predictive power of each modality (EEG, EDA or BVP) as well as of their fusion in a feature-level basis, in both single-class and multi-class scenarios (see Table~\ref{tab:classexps}). We exploited the effect of the number of estimators $[150, 300, 600]$ as well as of the maximum number of features $[ .5, 1, 2] *\sqrt{\text{NumberOfFeatures}}$ by means of grid search parameter estimation with 5-fold cross-validation (see also below). Overall, the optimum number of estimators was 300 and the maximum number of features was set equal to the total number of features for each experiment. The relative rank (i.e. depth) of each feature was estimated based on the ``Gini"  impurity function to assess the relative importance of that feature to the predictability of the target variable \cite{breiman2001}. 

All data from all times each participant walked a route were employed for the analysis. While overall familiarity might have gradually increased, individual environments still retained a dynamic complexity due to “new” stressors such as people coming from the opposite direction (outdoor) or out of the elevator (indoor), classroom doors opening (indoor), bicycles or cars being parked in different spots (outdoor), or chairs and tables being displaced (indoor). 
With regard to the outdoor collected dataset, there were 10,340 data points in total; the eight classes were significantly imbalanced, ranging from 3,278 (most frequent) to 460 (least frequent) data points. The indoor dataset comprised 6,412 data points in total; again the five classes were imbalanced ranging from 1,964 (most frequent) to 570 (less frequent) data points.

Sequential data points were split randomly in training and testing subsets (which, as a result, no longer contain sequential points). We trained one model for each of the single-class cases and one for the multi-class experiment following a 5-fold cross-validation scheme, where the 80\% of the data points were used for training and the 20\% for testing, with data shuffling in order to avoid dependencies in consecutive data points. The best model is chosen as the one that maximized the weighted area under receiver operating characteristic (AUROC) statistic, taking into account the lack of balance between the class labels.

\begin{table}[!t]
\centering
\caption{Classification Schemes}
\renewcommand{\arraystretch}{1.1} 
\begin{tabularx}{\columnwidth}{@{\extracolsep{\fill}}l >{\raggedright\arraybackslash}p{7.5cm}@{}}
\toprule
\textbf{Exp.} & \textbf{Description} \\
\midrule
\textbf{I}   &  Single-class classification using as predictors the unimodal features extracted from the EEG signals ($N = 198$).   \\ 
\textbf{II}  &  Single-class classification using as predictors the unimodal features extracted from the EDA signals ($N = 6$).  \\ 
\textbf{III} &  Single-class classification using feature-level multimodal fusion of raw EDA, BVP, and HR signals ($N = 3$).  \\ 
\textbf{IV}  &  Single-class classification using feature-level multimodal fusion of EEG and EDA features ($N = 204$).   \\ 
\textbf{V}  &  Single-class classification using feature-level multimodal fusion of EEG, EDA, BVP, and HR features ($N = 206$).   \\ 
\textbf{VI} & Multi-class classification using as predictors the unimodal features extracted from the EEG signals ($N = 198$) \\
\textbf{VII} & Multi-class classification using as predictors the unimodal features extracted from the EDA signals ($N = 6$) \\
\textbf{VIII} & Multi-class classification using feature-level multimodal fusion of raw EDA, BVP, and HR signals ($N = 3$) \\ 
\textbf{IX} & Multi-class classification using feature-level multi-modal fusion of EEG and EDA features ($N = 204$). \\
\textbf{X} & Multi-class classification using feature-level multimodal fusion of EEG, EDA, BVP, and HR features ($N = 206$) \\
\bottomrule
\end{tabularx}
\label{tab:classexps}
\end{table}

\subsection{Linear Mixed Model Analysis}\label{sec:lma}

We examined the role of vision impairment in the perception of environmental and situational stressors with a statistical analysis of features that emerged as the most predictive in the multimodal classification experiments. A linear mixed model method was used, which performs a regression-like analysis while controlling for random variance caused by differences in factors such as participant and electrode \cite{laird1982}. 
Fixed factors examined in the analyses included environment and vision. For the latter, three categories were considered: normal (visual acuity greater than 30\%), severe impairment (visual acuity less than 10\% but greater than 2\%), and blind (visual acuity less than 2\%). When fitting EDA and BVP data, a random intercept for each participant was added. When fitting EEG data, a further random intercept for electrode was included. Type III Wald $F$-tests were used to test the significance of the fixed factors and their interaction \cite{fox2011}. Pairwise comparisons of group means were carried out with $t$-tests, using Bonferroni-adjusted $p$-values where appropriate.

We previously described that EEG signals collected from visually impaired participants were baseline-normalized by subtracting the mean of resting state registrations obtained in a series of laboratory experiments, whereas for those of normally sighted participants the mean signal value was subtracted instead (Sec.~\ref{sec:features}). To facilitate comparison between the two groups, we re-baseline-normalized the EEG data of the VIP that walked the outdoor route to match those of the normally sighted. 

\section{Results and Discussion}

Due to temporary dysfunctions of the equipment, incomplete data recorded from two visually impaired participants during the first walk of the outdoor route and from two other participants during the first walk of the indoor route were discarded. 

\subsection{Stressful Environment Prediction}

The average weighted AUROC and standard deviation over five folds for all outdoor scene classification experiments are reported in Table \ref{tab:auroc_outdoor}. In the one-versus-all scenario (Exp.~I--V), the EEG and EDA modalities (Exp.~I and II, respectively) were both predictive of the distinct scenes (classes) and with highly similar performance. Fusing the two modalities (Exp.~IV) gave marginally improved results. The fusion of the EDA and BVP modalities (Exp.~III) boosted the performance of the classifier compared to using only EDA or EEG features or, to a lesser extent, both modalities. Combining features from all three modalities achieved almost perfect accuracy across all scenes. Similar trends were observed for the multi-class classification experiments (Exp.~VI--X). Including the BVP modality (Exp.~VIII and X) appeared to considerably improve performance over considering only EEG, EDA, or their fusion.

Table \ref{tab:auroc_indoor} summarizes the mean weighted AUROC and standard deviation over five folds for all indoor scene classification experiments. Overall performance for the indoor scenes was quite satisfactory, but not as high as for the outdoor scenes. In the one-versus-all scenario (Exp.~I--V), the EEG and EDA modalities (Exp.~I and II, respectively) were both predictive of the distinct scenes, with EEG performing considerably better than EDA. Fusing the two modalities (Exp.~IV) resulted in considerably better results, particularly of the elevator class, the detection of which improved substantially compared to when using only EDA features. Combining EDA with BVP (Exp.~III) achieved substantially better accuracy than using only EDA. Adding EEG (Exp.~V) only improved results for two classes (elevator and stairs). In the multi-class scenario, EEG (Exp.~VI) performed better than EDA (Exp.~VIII). Their fusion (Exp.~IX) marginally increased accuracy. Combining EDA with BVP gave the best outcome, softly outperforming the fusion of all three modalities (Exp.~X). 

\begin{figure}[!b]
\centering
\includegraphics[width=\columnwidth]{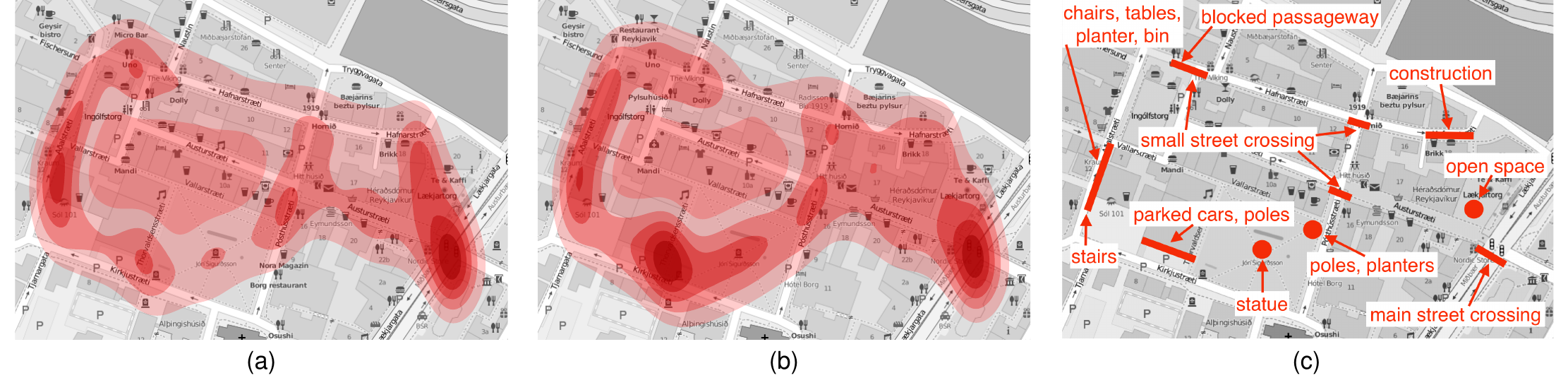}
\caption{One-against-all weighted ROC curves for the multi-class multimodal classification (Exp.~X in Fig.~\ref{tab:classexps}).}
\label{fig:roc}
\end{figure}

As a means of assessing the qualitative performance of the multi-class multimodal fusion model (Exp.~X), Fig.~\ref{fig:roc} shows the weighted ROC curves for each outdoor and indoor scene in a one-against-all binary scenario. In both cases (outdoor and indoor), the trained model was able to learn all different scenes equally well, providing proof of the stability of the multimodal approach. 

\begin{table*}
\centering
\caption{Indoor Scene Classification Average Weighted AUROC and Standard Deviation Over 5 Folds}
\renewcommand{\arraystretch}{1.1}
\begin{tabular*}{\textwidth}{@{\extracolsep{\fill}}lccccc@{}}
\toprule
\textbf{Environment} & \textbf{EEG} & \textbf{EDA} & \textbf{\{EDA,BVP\}} & \textbf{\{EEG,EDA\}} & \textbf{\{EEG,EDA,BVP\}}\\
\midrule
\emph{\textbf{Single-Class Classification}} & \textbf{Exp.~I} & \textbf{Exp.~II} & \textbf{Exp.~III} & \textbf{Exp.~IV} & \textbf{Exp.~V}\\
\cmidrule(lr){2-3} \cmidrule(l){4-6} 
Door (A)             & 76 (0.6)  & 76 (1.2)  & 86 (1.1)  & 79 (0.9) & 86 (1.0)\\  
Elevator (B)       & 87 (0.8)  & 79 (0.3)  & 88 (0.8)  & 89 (0.8) & 93 (0.8)\\  
Corridor (C)       & 72 (1.6)  & 66 (1.9)  & 79 (1.3)  & 75 (0.9) & 76 (0.8)\\  
Open space (D) & 76 (0.8)  & 67 (1.0)  & 85 (0.9)  & 77 (1.2) & 85 (0.7)\\  
Stairs (E)           & 78 (1.2)  & 69 (1.1)  & 84 (1.0)  & 82 (1.2) & 92 (0.4)\\
\midrule
\emph{\textbf{Multi-Class Classification}} & \textbf{Exp.~VI} & \textbf{Exp.~VII} & \textbf{Exp.~VIII} & \textbf{Exp.~IX} & \textbf{Exp.~X} \\
\cmidrule(lr){2-3} \cmidrule(l){4-6} 
All indoor & 78 (0.4) & 71 (0.6) & 87 (0.7) & 81 (0.5) & 84 (0.5)\\
\bottomrule
\end{tabular*}
\label{tab:auroc_indoor}
\end{table*}

Feature importances were estimated for all multi-class experiments. The most predictive ones appeared always with the highest ranks. In the outdoor scene classification, mean tonic EDA (TM) emerged as the most predictive feature in Exp.~VII (EDA features only), Exp.~IX (fusion of EEG and EDA features), and Exp.~X (fusion of EEG, EDA, and BVP features). The same results were obtained for the indoor scene experiments, except for Exp.~X where heart rate (HR) performed marginally better than TM. The emergence of the latter as the most predictive feature in the present experiments confirms previous psychophysiological findings showing skin conductance tonic level to be a highly relevant index of stress-induced physiological arousal.

Among the eight most predictive features resulting from the \{EEG,EDA,BVP\} fusion models were the EDA and HR signals registered by the Empatica E4 wristband, which were used ``as is'' (but min-max scaled to reduce inter-individual variation, see Sec.~\ref{sec:features}). This result illustrates the big potential of using existing state-of-the-art sensors such as the E4 for real-time prediction of human affective states from peripheral physiological signals during less controlled experimental conditions, which could be employed to design emotionally intelligent mobility aids for visually impaired travelers.

Predictions involving EEG features, alone or fused with other modalities, were dominated by changes in spectral power (i.e., ERD/ERS, see Sec.~\ref{sec:features}) in the delta band of the F3/4, T7/8, P7/8, and, to a lesser extent, FC5/6 and O1/2 channels. Although real-time EEG acquisition may be subject to very noisy signals, this finding is in line with neuropsychophysiological evidence (a) reporting increased delta activity during mental tasks as a result of attention to internal processing \cite{harmony1996} and (b) suggesting that the 10-20 system locations F3/4, F7/8 and T7/8 may be suitable enough to monitor brain activity under cognitive-emotional stress \cite{zheng2015}. 

EDA and BVP features were generally better in predicting stressful urban scenes, whereas the EEG modality performed better in indoor scene classification. This might have resulted from differences in the how the outdoor and indoor routes were designed. Whereas the former focused mainly on passive walking, the latter involved active wayfinding, for example, turning and going towards the door, finding where stairs begin, negotiating orientation while climbing down the spiral stairs and after exiting the elevator. Therefore, whereas EDA and BVP features reflected the more general stressful situations across urban settings, EEG features traced changes in cognitive load during specific indoor wayfinding tasks. This difference between the two studies proves further evidence of the complementary nature of the three modalities in assessing human affective states and thus supports the need for multimodal approaches to stress detection in visually impaired mobility.

Person-specific effects on the biomarkers were also assessed through performing a leave-one-participant-out cross-validation. Overall results were greatly affected, dropping in some cases as much as 50\% weighted AUROC. This is not surprising given the nature of human rain and physiological electrical signals. Individual reactions should be considered when designing automatic affective state recognition for navigation aid systems. The proposed models are a viable fit for personalized systems, where after a short period of user-specific training almost perfect accuracies can be achieved. 

\begin{figure*}[!t]
\centering
\includegraphics[width=\textwidth]{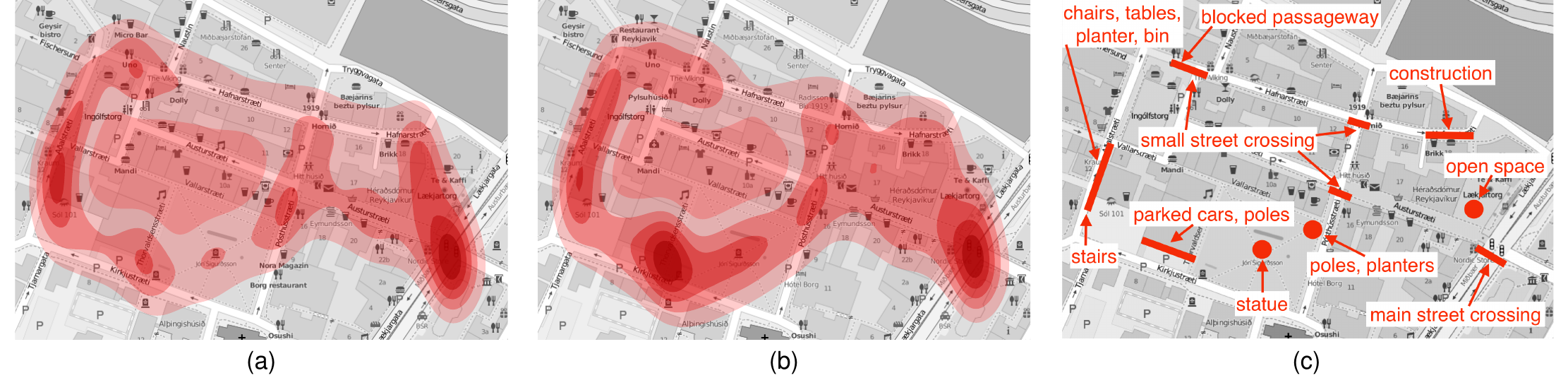}
\caption{Outdoor route geographic density distributions of (a) mean tonic EDA (TM) and (b) number of spontaneous skin conductance peaks (SCRs). The darker the color is, the higher the density of the distribution is. See text for details on spatial density estimation. (c) Annotated obstacles and situations along the route (see upper panel of Table \ref{tab:environments}).}
\label{fig:densout}
\end{figure*}

\begin{figure}[!b]
\centering
\includegraphics[width=0.9\columnwidth]{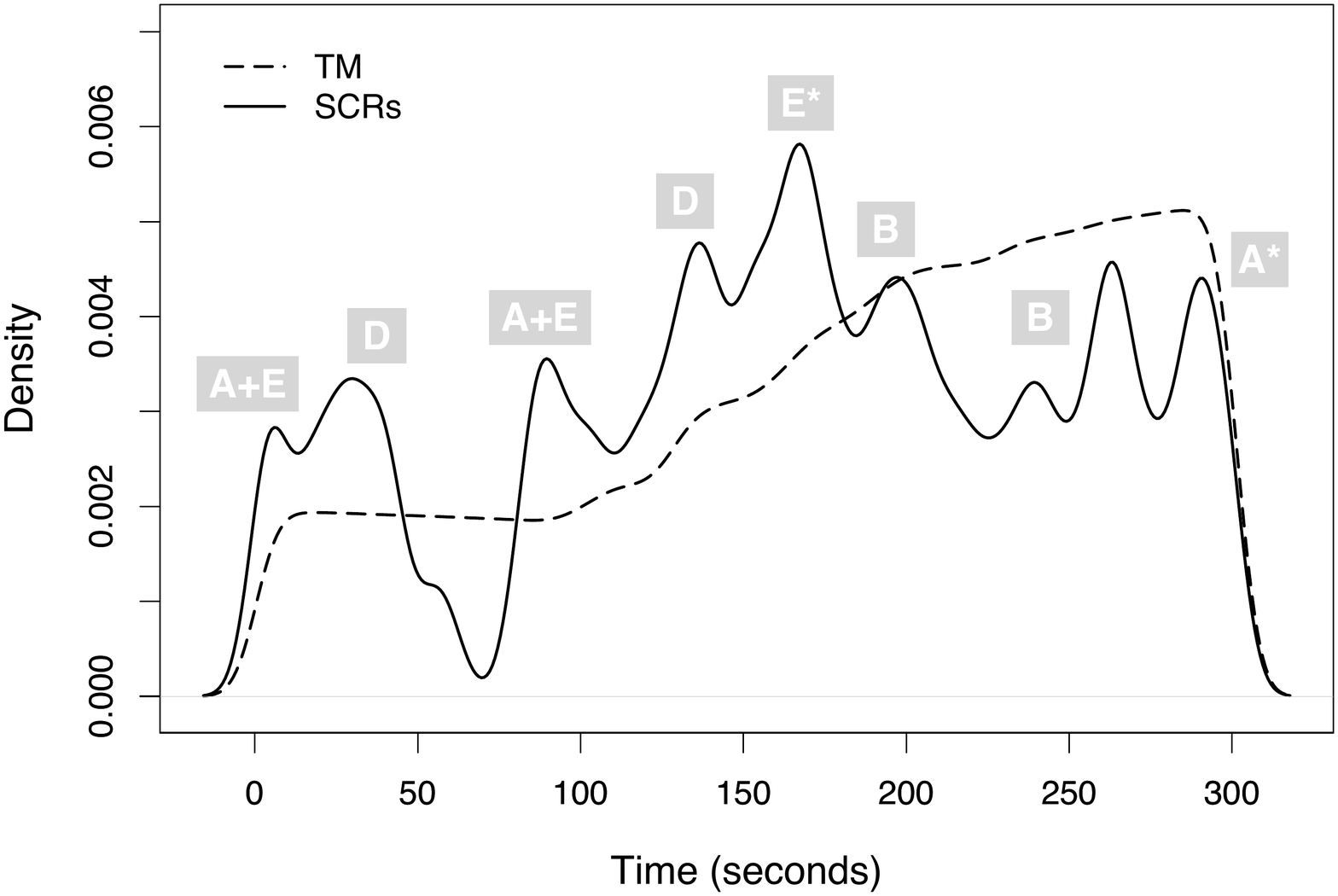}
\caption{Indoor route temporal density distributions of mean tonic EDA (TM) and number of spontaneous skin conductance peaks (SCRs). See text for details on temporal density estimation. Letters depict the different indoor scenes reported in Table \ref{tab:environments} (lower panel), with A$^*$ indicating the rotating door and E$^*$ the large spiral stairs.}
\label{fig:densind}
\end{figure}

\subsection{Visualizing Biomarker Density Distributions}

To better understand how mean tonic EDA (TM) relates to environmental and situational factors, as well as the intensity of the cognitive and emotional response it expresses, its geographical (outdoor route) and temporal (indoor route) density distributions were assessed by means of weighted kernel density estimation \cite{wand1995}. We contrasted TM density with that of spontaneous skin conductance peaks (SCRs), the least predictive EDA feature. While the former reflects long-term tonic arousal, SCRs capture transitory increases in tonic skin conductance. 

Let $ \{ \mathbf{x}_1, \mathbf{x}_2, \ldots, \mathbf{x}_n \} $ be an independent random sample drawn from some distribution with density function $f(\mathbf{x})$ defined on $\mathbb{R}^d$. The weighted kernel density estimate of $f$ is defined as
\[
\hat{f}_H(\mathbf{x}) = \frac{1}{n} \displaystyle\sum_{i=1}^{n}{ w(\mathbf{x}_i,\mathbf{w})\: K_H (\mathbf{x}-\mathbf{x}_i)}
\]
where $K$ is a kernel function, $H > 0$ is a symmetric $ d \times d $ matrix which controls the bandwidth (or smoothing) of the estimate, $K_H(\mathbf{x}) = |H|^{\sfrac{-1}{2}} K(H^{\sfrac{-1}{2}\:} \mathbf{x})$, and $w$ is a function weighting each data point in the sample with a value from $\mathbf{w} \in \mathbb{R}^m,\: m \leq d $. A popular choice for $K$ is the Gaussian (or normal) kernel, which was also applied here. 

For the outdoor scenes, feature values were assigned to pairs of latitude and longitude coordinates based on recorded timestamps. Using values as weights ($w$ with $m = 1$) for GPS points ($\mathbf{x}$ with $d = 2$) and a bandwidth of $H(\mathbf{x}) = 0.0008$ helped estimate the feature-weighted density of GPS points on a $500 \times 500$  grid, and based on this generate a density distribution contour plot for each participant (VIP only). Figure \ref{fig:densout} shows the resulting TM and SCRs contours (left and middle maps, respectively) aggregated for all participants and walks and plotted on top of an OSM map (the darker the color, the higher the density of the distribution). 

Increased stress-elicited arousal along the different urban scenes of the route is immediately observed when the VIP had to cross a main road (scene F, see Fig.~\ref{fig:route_out} and upper panel of Table \ref{tab:environments}), pass along parked cars in a narrow alley after the urban park (C), walk up and down stairs (E), or pass through a narrow area between construction works (H). These arousal ``hotspots'' are in full agreement with the scenes reported as stressful by the participants themselves at the end of the study. Furthermore, they coincide with the presence of certain obstacles and situations that can be less or more stressful in visually impaired mobility, and which were taken into account when designing the outdoor route (Fig.~\ref{fig:densout}, right panel). 


To visualize density distribution along the indoor route, feature values were assigned to 1-second steps from the start point based on recorded timestamps. Due to different walking speeds and behaviors, individual walk times varied between participants and trials, ranging from 4 to 8 minutes with an average length of 5 minutes. To temporary align all features so that same times corresponded to same environments we performed dynamic time warping \cite{giorgino2009}, postulating that a certain environment induced similar biomarker patterns. Each feature vector was warped to a reference vector that was 300 seconds (5 minutes) long. Using warped feature values as weights ($w$ with $m = 1$) for 1-second time steps ($\mathbf{x}$ with $d = 1$) and a bandwidth of $H(\mathbf{x}) = 5.59$ helped estimate the feature-weighted density of time points (temporal distances) on a $400$-point grid, and based on this generate a density function across all participants and walks. 

Figure \ref{fig:densind} shows the resulting density distributions for the TM and SCRs features plotted together and annotated with the different indoor scenes reported in the lower panel of Table \ref{tab:environments}. Tonic skin conductance appears to gradually increase towards the second half of the walk, which is where the most stressful parts of the route were according to reports from all participants at the end of the study. As expected, the estimated density of SCRs generally followed the trend of TM while allowing to observe localized rises in arousal. These suggest the presence of a higher number of instantaneous stressors, for example, when safely entering/exiting an elevator or passing through a rotating door or walking up/down stairs while others try to do the same, or maintaining direction amidst loud noises and people moving in an open space indoors.   

            

\begin{table*}[!t]
\centering
\caption{Linear Mixed Model Type III Wald Tests Comparing Categories of Visual Impairment}
\label{tab:Ftests1}
\renewcommand{\arraystretch}{1.1}
\begin{tabularx}{\textwidth}{@{\extracolsep{\fill}}lrrrrrr@{}}
\toprule
& df~~~~ & $F$~~ & $p$~~~ & df~~ & $F$~~ & $p$~~~ \\
\midrule 
& \multicolumn{3}{c}{\textbf{Outdoor}} 
& \multicolumn{3}{c}{\textbf{Indoor}}     \\
\midrule 
& \multicolumn{6}{c}{\textbf{TM}}  \\
\midrule
Intercept (I) & 1, 6.00    & 24.47 & 0.003    & 1, 7.03 & 29.27 & $<$ 0.001 \\
Vision (V)    & 1, 6.00    & 0.72 & 0.430     & 1, 7.03 & 0.96 & 0.360 \\
Scene (S)     & 7, 89.01  & 0.76 & 0.621      & 4, 108.00 & 1.25 & 0.294 \\
V $\times$ S  & 7, 89.01  & 0.33 & 0.937      & 4, 108.00 & 0.02 & 0.999 \\
\midrule
& \multicolumn{6}{c}{\textbf{HR}}  \\
\midrule
I             & 1, 6.00 & 33.22 & 0.001     & 1, 7.06 & 76.41 & $<$ 0.001 \\
V             & 1, 6.00 & 1.16 & 0.323      & 1, 7.06 & 0.04 & 0.848 \\
S             & 7, 89.01 & 4.83 & $<$ 0.001 & 4, 108.01 & 0.79 & 0.533 \\
V $\times$ S  & 7, 89.01 & 2.58 & 0.018     & 4, 108.01 & 0.24 & 0.914 \\
\midrule
& \multicolumn{6}{c}{\textbf{ERD/ERS, delta}}  \\
\midrule
I & 1, 6.06 & 23.55 & 0.003                 & 1, 9.55 & 100.24 & $<$ 0.001 \\
V & 1, 6.08 & 0.66 & 0.447                  & 1, 7.51 & 2.48 & 0.157 \\
S & 7, 639.20 & 3.92 & $<$ 0.001            & 4, 832.31 & 16.87 & $<$ 0.001 \\
V $\times$ S & 7, 639.77 & 5.63 & $<$ 0.001 & 4, 831.92 & 0.52 & 0.719 \\
\midrule
& \multicolumn{6}{c}{\textbf{ERD/ERS, alpha-2}} \\
\midrule
I & 1, 7.97 & 1027.09 & $<$ 0.001 & 1, 9.74 & 903.41 & $<$ 0.001 \\
V & 1, 6.05 & 7.38 & 0.035 & 1, 7.37 & 1.23 & 0.302 \\
S & 7, 833.86 & 5.39 & $<$ 0.001 & 4, 888.74 & 20.92 & $<$ 0.001 \\
V $\times$ S & 7, 833.90 & 7.30 & $<$ 0.001 & 4, 887.90 & 1.16 & 0.325 \\
\midrule
& \multicolumn{6}{c}{\textbf{FAI}} \\
\midrule
I & 1, 6.00 & 0.17 & 0.699                & 1, 7.04 & 0.12 & 0.741 \\
V & 1, 6.00 & 1.45  & 0.274               & 1, 7.04 & 3.97  & 0.087     \\
S  & 7, 89.01 & 0.07  & 0.999             & 4, 108.00 & 1.15  & 0.335     \\
V $\times$ S  & 7, 89.01 & 0.10  & 0.998  & 4, 108.00 & 0.15  & 0.964     \\
\bottomrule
\end{tabularx}
\end{table*}

\subsection{Linear Mixed Model Analysis}
            
Mean tonic EDA (TM) and heart rate (HR), the two most predictive biomarkers, were first analyzed. We then examined ERD/ERS in delta frequencies, an EEG feature that further dominated predictions. In addition, because higher alpha desynchronization has been consistently associated with increased cognitive load \cite{Antonenko2010} and asymmetry in frontal alpha activity has been found to relate to the positive/negative disposition and engagement \cite{harmonjones2010}, ERD/ERS and FAI values in the upper alpha band (alpha-2) were also analyzed. 
Before averaging across conditions, a logarithmic transformation of single-condition feature values was applied to improve their distributional characteristics, except for FAI as its definition involves such a transformation already. 

Type III Wald $F$-tests comparing two categories of visual impairment (severe versus blind, see Sec.~\ref{sec:lma}) are displayed in Table \ref{tab:Ftests1}. Vision alone was only a significant predictor of upper alpha ERD/ERS in the outdoor route and a marginally significant predictor of FAI in the indoor route, although the interaction of vision and scene was significantly influential for both delta and upper alpha ERD/ERS as well as for HR in the outdoor route.\footnote{By ''marginally significant'' we denote a value that is close enough to the typical threshold of $p = 0.05$ to be ruled out as not significant.} The scene alone had a significant effect on delta and alpha ERD/ERS in both outdoor and indoor models, as well as on HR in the outdoor model.  

Post-hoc paired samples $t$-tests showed that HR was significantly lower for severely impaired than for blind individuals when crossing a major intersection [outdoor environment E, $t(13.21)=-2.45, p=0.029$] and when walking in a shopping street [outdoor scene A, $t(13.21)=-2.39, p=0.033$]. ERD/ERS in the delta band varied significantly between the two VIP groups for the outdoor environments F [crossing main road with traffic lights, Severe $>$ Blind, $t(31.29)=2.43, p=0.021$], G [crossing small street without traffic lights, Severe $<$ Blind, $t(13.54)=-2.99, p=0.010$], and E [open space, Severe $<$ Blind, $t(26.80)=-2.20, p=0.037$], and marginally significantly for outdoor scene H [construction alley, Severe $<$ Blind, $t(14.14)=-1.80, p=0.094$]. For the indoor environments, delta EDR/ERS was only marginally significantly higher for blind than for severely impaired individuals when navigating through automated moving doors [$t(28.00)=1.95, p=0.062$]. ERD/ERS in the upper alpha band was significantly lower for severely impaired than for blind individuals for the outdoor environments A [$t(13.52)=-2.35, p=0.035$], B [small street, $t(20.51)=-4.13, p=0.001$], E [$t(15.85)=-2.14, p=0.048$], G [$t(25.92)=-4.20, p<0.001$], and H [$t(24.07)=-2.47, p=0.021$]. 
No significant within-scene differences emerged between the two VIP groups in indoor environments. Finally, FAI was only marginally significantly higher for severely impaired than for blind individuals when walking along a narrow corridor [indoor scene C, $t(15.63)=1.95, p=0.069$].


\begin{table}[!t]
\centering
\caption{Linear Mixed Model Type III Wald Tests Comparing Impaired to Normal Vision (Outdoor Study Only)}
\label{tab:Ftests2}
\renewcommand{\arraystretch}{1.1}
\begin{tabularx}{\columnwidth}{@{\extracolsep{\fill}}lrrr@{}}
\toprule
& df~~~~ & $F$~~ & $p$~~~ \\
\midrule
& \multicolumn{3}{c}{\textbf{TM}} \\
\midrule
Intercept (I) & 1, 6.95    & 27.58 & 0.001     \\
Vision (V)    & 2, 6.99    & 0.56 & 0.594     \\
Scene (S)     & 7, 112.00  & 4.92 & $<$ 0.001 \\
V $\times$ S  & 14, 112.00 & 1.63 & 0.082      \\
\midrule
& \multicolumn{3}{c}{\textbf{HR}}    \\
I & 1, 6.91 & 31.51 & $<$ 0.001 \\
V & 2, 6.93 & 1.49 & 0.291 \\
S & 7, 112.01 & 5.64 & $<$ 0.001 \\
V $\times$ S  & 14, 112.01 & 1.42 & 0.154 \\
\midrule
& \multicolumn{3}{c}{\textbf{ERD/ERS, delta}} \\
\midrule
I & 1, 7.08 & 13.32 & 0.008   \\
V & 2, 7.01 & 0.53 & 0.610   \\
S & 7, 831.65 & 6.35 & $<$ 0.001   \\
V $\times$ S & 14, 831.47 & 4.17 & $<$ 0.001  \\
\midrule
& \multicolumn{3}{c@{}}{\textbf{ERD/ERS, alpha-2}}     \\
I  & 1, 7.46 & 514.59 & $<$ 0.001  \\
V  &  2, 7.00 & 1.53 & 0.281  \\
S  &  7, 1069.18 & 5.88 & $<$ 0.001  \\
V $\times$ S &  14, 1069.05 & 5.03 & $<$ 0.001  \\
\midrule
& \multicolumn{3}{c}{\textbf{FAI}} \\
\midrule
I & 1, 6.92 & 0.42 & 0.540  \\
V & 12, 6.94 & 0.87  & 0.460        \\
S  & 7, 112.01 & 0.14 & 0.995          \\
V $\times$ S  & 7, 112.01 & 0.10  & 1.000      \\
\bottomrule
\end{tabularx}
\end{table}

Table \ref{tab:Ftests2} reports type III Wald $F$-tests comparing impaired to normal vision (outdoor route only), specifically normal vision to severe visual impairment on the one hand, and normal vision to blindness on the other. No significant differences emerged between the three categories of vision for the examined biomarkers. Scene was a significant predictor in all models but FAI. The two factors appeared to influence each other significantly for delta and alpha ERD/ERS, and nearly significantly for TM.

Post-hoc paired samples $t$-tests showed no significant within-scene differences between pairs of the three vision groups for neither of the tested biomarkers. The significant interaction between vision and scene in the TM and ERD/ERS models was a result of differences between severely impaired and blind individuals rather than between these and normally sighted persons. Accordingly, we merged the Severe and Blind participants into a single VIP group and run new $t$-tests against the group of normally sighted. Still, no significant within-scene differences emerged between normal and impaired vision. When averaging across scenes, no significant differences emerged between normal and impaired vision either. This result could have two different origins considering the circumstances related to the particular study. First, the VIP who took part are super-achievers: they have a job or attend college, and travel alone outdoors on an almost daily basis. Second, they were accompanied by their familiar O\&M instructor to help make them feel comfortable and safe. These experimental factors may have prevented the elicitation of increased psychological stress.  

\section{Conclusions}

Mobility aids for visually impaired people should be capable of implicitly adapting not only to changing environments but also to shifts in the affective state of the user in relation to different environmental and situational factors. To this end, this paper presents a framework for real-time automatic assessment of the cognitive-emotional experience of VIP while navigating in unfamiliar outdoor and indoor environments, based on ambulatory monitoring and fusion of brain and peripheral biosignal data. Different multimodal fusion scenarios were compared, aiming to address the robustness of the model as well as emerging differences in the perception and interaction of VIP with their surroundings. 

The consistently high prediction rates in the multimodal classification experiments (81--93\% weighted AUROC, Table \ref{tab:auroc_outdoor} and \ref{tab:auroc_indoor}) are very encouraging of the proposed approach. Even if the chosen city and building sites did not represent all possible different outdoor and indoor environments and situations in terms of complexity and difficulty, the charted routes were designed so as to combine most of the mobility challenges faced by VIP. Indeed, the most predictive biomarkers indicated spaces and situations as stressful and cognitively demanding ''hotspots'' in perfect agreement with the self-reported experience of the visually impaired participants (Fig.~\ref{fig:densout} and \ref{fig:densind}).

Reported findings, despite being promising, should be considered with caution due to the limited number of participants, which did not allow for an in-depth analysis of specific stressors in each category of vision impairment. Furthermore, the well-established Emotiv EPOC\verb!+! EEG headset has certain limitations with respect to the quality of the recorded signal during experiments involving physical activity ``in the wild'' such as those presented in this paper. The number of the provided electrodes is limited and hence the EEG markers discussed in this paper are meant to provide only insights on the most predictive features and their connection to specific tasks and conditions. 

A rich multimodal dataset has been collected, which will be made openly available in order to maximize the impact of the work and encourage further investigations. Future steps of the present study include refining the predictive model through exploring novel multimodal biosignal features and comparing different classifiers. Such findings hopefully pave the way to emotionally intelligent mobile technologies that take the concept of navigation one step further, accounting not only for the shortest path but also for the most effortless, least stressful and safest one.

\ifCLASSOPTIONcompsoc
  \section*{Acknowledgments}
\else
  \section*{Acknowledgment}
\fi

The research leading to these results has received funding from the European Union's Horizon 2020 Research and Innovation program under grant agreement No 643636 ''Sound of Vision'' and the Alexander von Humboldt Foundation through a Humboldt Research Fellowship awarded to CS. KK acknowledges support from the ``Lagrange Project'' of the ISI Foundation funded by the Fondazione CRT.
The authors wish to thank the administration and O\&M instructors at the National Institute for the Blind, Visually Impaired, and Deafblind in Iceland for their valuable input and generous assistance, as well as the visually impaired individuals who took part in the study for their time and patience.

\ifCLASSOPTIONcaptionsoff
  \newpage
\fi



\bibliographystyle{IEEEtran}
\bibliography{citations}
\FloatBarrier

\begin{IEEEbiography}[{\includegraphics[width=1in,height=1.25in,clip,keepaspectratio]{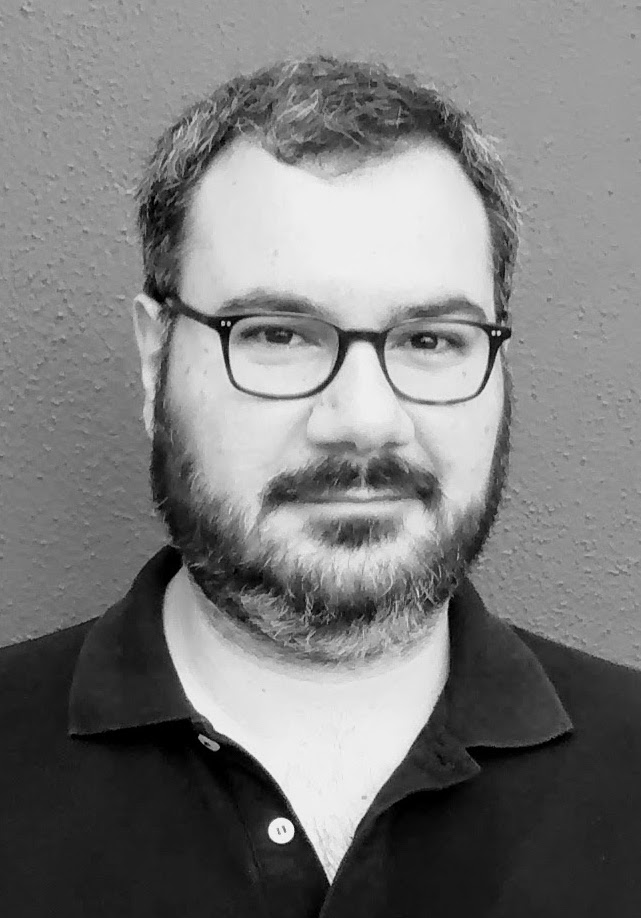}}]{Charalampos Saitis} holds an MA in Sonic Arts from Queen's University Belfast and a PhD in Music Technology from McGill University. He is currently Humboldt Research Fellow at the Technical University of Berlin. His work focuses on psychoacoustics and auditory semantics, particularly on approaches relating to embodied cognition and crossmodal processing. Other research interests include haptics in music and affective biosignal computing for cognitive load assessment.
\end{IEEEbiography}

\begin{IEEEbiography}[{\includegraphics[width=1in,height=1.25in,clip,keepaspectratio]{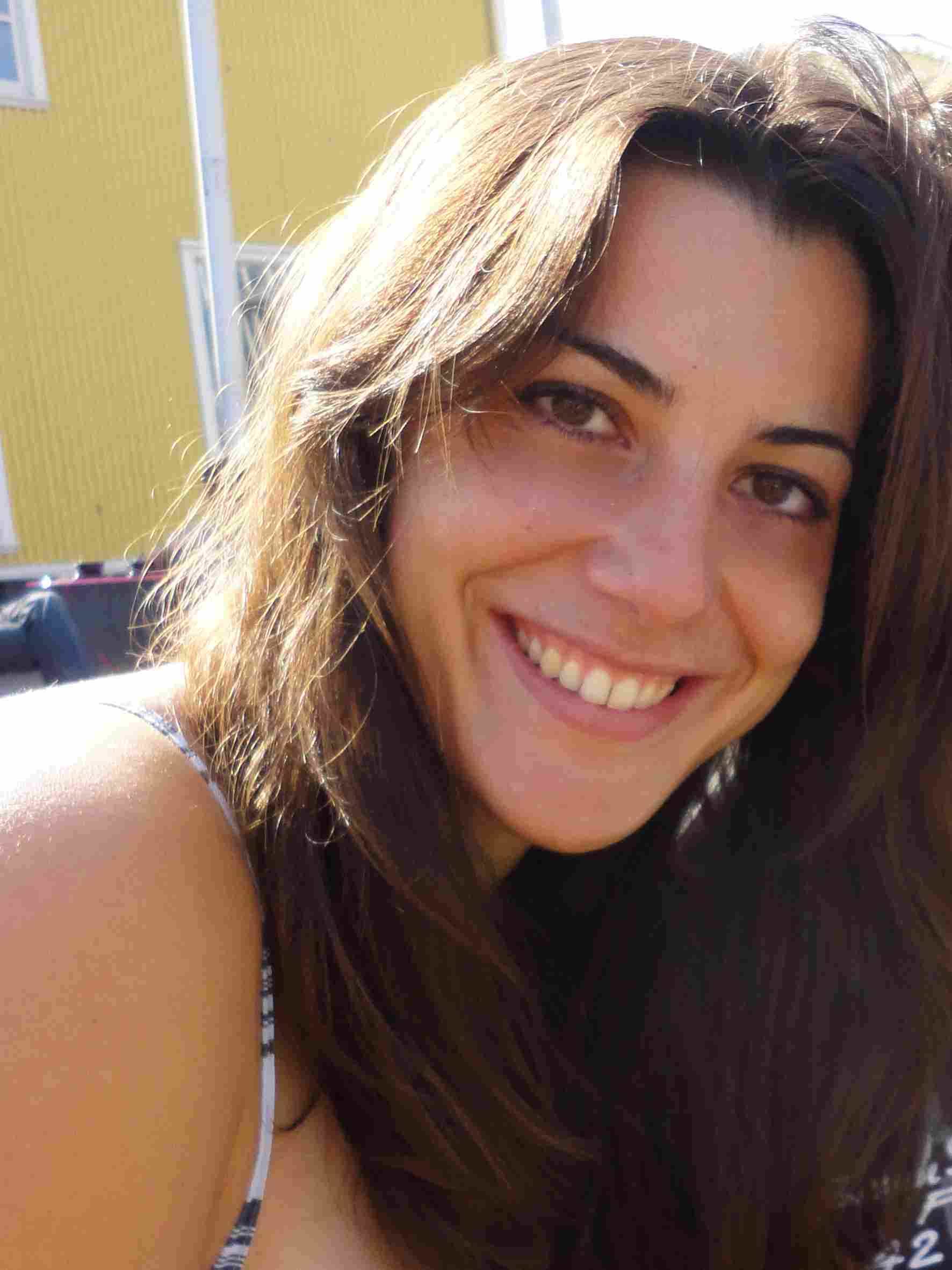}}]{Kyriaki Kalimeri}
holds a PhD in Brain and Cognitive Science from the University of Trento and a Diploma in Electrical and Computer Engineering from the Technical University of Crete. Currently, she is Researcher at the ISI Foundation. Previously she was research assistant at the Fondazione Bruno Kessler and visiting researcher at the Human Dynamics Group of the MIT Media Lab. Her research lies at the intersection of social science and engineering. She is particularly interested in employing machine learning techniques for the prediction of psychometric and affective profiles from physiological, smartphone, and social media data. 
\end{IEEEbiography}

\end{document}